\documentclass[a4paper,11pt]{article}

\usepackage{jcappub}
\usepackage[utf8]{inputenc}
\usepackage{amsmath,amssymb}
\usepackage{graphicx}
\usepackage{multicol}
\usepackage{hyperref}
\usepackage{url}
\usepackage{physics}
\usepackage{lipsum}
\usepackage{caption}
\usepackage{subcaption}
\usepackage{accents}
\usepackage{caption}
\usepackage{multirow}
\usepackage{float}
\usepackage{soul}
\usepackage{comment}
\usepackage{graphicx}
\usepackage{tabularx}
\usepackage{xcolor}
\usepackage[export]{adjustbox}
\usepackage{bigfoot}

\usepackage{cases}
\usepackage{amsfonts}
\usepackage{dcolumn}
\usepackage{bm}
\usepackage{bbm}
\usepackage{array}
\usepackage{wasysym}

\usepackage[T1]{fontenc}
\usepackage{bm,tensor}
\usepackage{braket}
\usepackage{ifpdf}
\usepackage{color}
\usepackage{ulem}

\usepackage{orcidlink}
\usepackage{array}
\usepackage{psfrag,fancyhdr,epsfig}
\usepackage{hyperref}

\usepackage{physics}
\usepackage{bigints}
\usepackage{verbatim}
\usepackage{booktabs}

\usepackage[mathscr]{eucal}

\newcommand{\m}{m_{\text{P}}}

\title{\bf Evading the BBN bound with a soft stiff period}

\author[a]{Lucy Brissenden,} 
\affiliation[a]{Consortium for Fundamental Physics, Physics Department,\\Lancaster University, Lancaster LA1 4YB, United Kingdom.}

\author[a]{Konstantinos Dimopoulos,}

\author[a,b]{Eemeli Tomberg}
\affiliation[b]{Cosmology, Universe and Relativity at Louvain (CURL), Institute of Mathematics and Physics, University of Louvain, 2 Chemin du Cyclotron, 1348 Louvain-la-Neuve, Belgium}

\abstract{
Cosmic inflation is the leading theory to explain early Universe history and structure formation. Non-oscillatory inflation is a class of models which can naturally introduce a post-inflationary stiff period of the Universe's evolution which boosts the signal of primordial gravitational waves (GWs), making it possible to observe them in forthcoming GW experiments. However, this pushes the GW energy density high enough to destabilise the process of Big Bang Nucleosynthesis (BBN). This problem can be overcome by ``softening'' the stiff period, so that the field is gradually tending towards freefall from a frozen start. Here, we consider a modified hybrid inflation model where the stiff period is driven by the waterfall field, allowing the barotropic parameter of the Universe to vary, so that it does not violate the $\Delta N_{\textnormal{eff}}$ constraint but produces a characteristic gravitational wave spectrum soon to be observable.}

\date{\today}

\begin{document}

\maketitle

\section{Introduction}
\label{sec:intro}
The most compelling solution to the horizon and flatness problems as well as the origin of the primordial curvature perturbation, necessary for structure formation to occur, is the theory of cosmic inflation \cite{Starobinsky:1980te,Guth:1980zm,Sato:1981qmu,Kazanas:1980tx}. The curvature perturbation generated during inflation is reflected onto the Cosmic Microwave Background (CMB) radiation \cite{Sachs:1967er}. Precision observations 
on the primordial anisotropy in the CMB have confirmed the inflationary predictions of acoustic peaks, which led to the collapse of the rival paradigm for structure formation, that of cosmic strings \cite{Perivolaropoulos:2005wa}. Another prediction of inflation is a stochastic gravitational wave background (SGWB), which is expected to be generated during inflation in a similar manner as the scalar curvature perturbation. 
Many consider the eventual observation of the SGWB to be an undeniable smoking gun for inflation \cite{Guzzetti:2016mkm}.

The basic inflation paradigm considers that, after inflation ends, the thermal bath of the hot big bang is promptly generated by the decay of the inflaton field (reheating) \cite{Lyth:2009imm}. In this case, the superhorizon modes of the tensor perturbations which were generated during inflation reenter the horizon during the radiation era. Then, the corresponding SGWB spectrum is flat: its spectral density $\Omega_{GW}$ does not depend on the frequency. Unfortunately, the amplitude of this flat GW spectrum is too small to be observed at present. However, there are a number of ways that observable (e.g. by LISA \cite{Bartolo:2016ami}) primordial GWs could be generated by inflation.

One prominent possibility is to consider a stiff post-inflationary period. Such a period is a natural occurrence in non-oscillatory (NO) models of inflation, where the inflaton field $\sigma$ is a runaway flat direction in field space.
In NO models, after inflation, the Universe continues to be dominated by the inflaton, but this time the field's kinetic energy density $K(\sigma)$ is much larger than its potential density $V(\sigma)$, meaning that its barotropic parameter (equation-of-state parameter) is approximately unity,
\begin{equation}
  w=\frac{K-V}{K+V}\approx 1\,.
 \label{wearly}
\end{equation}
Such a period is called kination \cite{Joyce:1997fc}. During kination, the field's equation of motion becomes oblivious to the scalar potential. We then say that the field is in freefall, subject only to Hubble friction.

\subsection{Effects of Stiff Dynamics on Primordial Gravitational Waves}
\label{subsec:stiffPGWs}

It has previously been demonstrated that a short period of kination might lead to observable primordial gravitational waves (GWs) \cite{Gouttenoire:2021jhk}, since it makes the spectral density parameter grow with frequency, $\Omega_{GW}\propto f$. However, there are severe restrictions on the length of a kination period \cite{
HaroCases:2020nsn}: in particular, the GW enhancement scales so steeply with the frequency that high-energy GWs threaten to disturb the delicate process of Big Bang Nucleosynthesis (BBN). This is encoded as a constraint on the integrated density of gravitational waves at BBN, given by
\begin{equation}
\int^{k_{\textnormal{end}}}_{k_{\textnormal{BBN}}}\dd \ln{k} \, h_{0}^{2} \, \Omega_{\textnormal{GW}}(k,\text{today}) \leq 5.6 \times 10^{-6} \, ,
\label{eq:dNeffconstraint}
\end{equation}
and is detailed further in Sec. \ref{subsec:constraints}. We refer to this bound henceforth as the $\Delta N_{\textnormal{eff}}$ constraint. Taking the $\Delta N_{\textnormal{eff}}$ constraint into account, the kination-generated peak in the GW spectrum cannot extend to frequencies low enough to be observable in the near future.

\begin{figure}
    \centering
    \includegraphics[width=\linewidth]{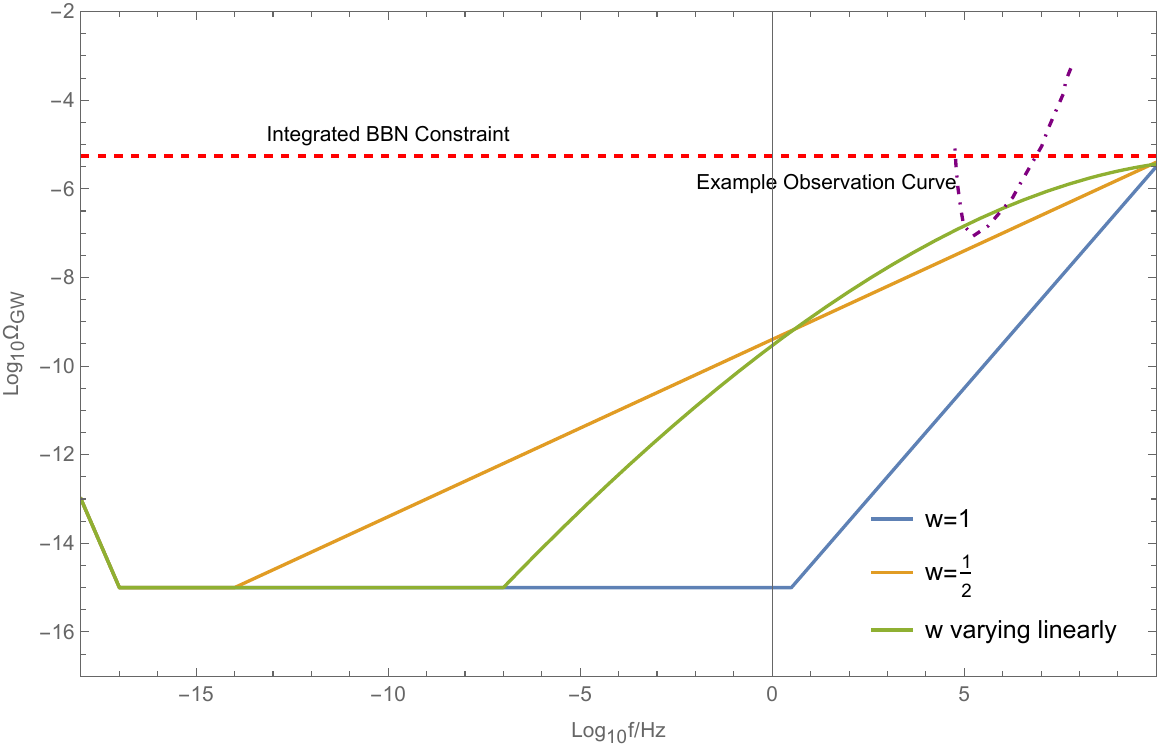}
    \caption{Conceptual sketch of how the $\Omega_{GW}$ spectrum behaves when subject to a scalar-field-driven period directly after (GUT-scale) inflation, with different barotropic parameters. Kination (blue) has the steepest proportionality and violates the $\Delta N_{\textnormal{eff}}$ constraint (shown here as the integrated BBN constraint) if it continues for too long, so even if kination does occur, it is unlikely to be observable.
    Unlike a stiff period of fixed barotropic parameter $w=\frac{1}{2}$ (orange), a period of varying barotropic parameter (green) is expected to form a curved peak; this allows a larger signal for $f\gtrsim 10$ Hz while still satisfying the BBN and $\Delta N_{\textnormal{eff}}$ bounds, making the spectrum more accessible to future observations. Here the variation in $w$ is modelled as linear in $\log(f)$ for illustrative purposes. The purple dashed line shows an example of how an instrumental observation curve can overlap with this slowly varying stiff period but not a stiff period of constant barotropic parameter. For $f \lesssim 10^{-17}$ Hz, the spectrum rises again due to the matter-dominated period at late times, and will eventually turn downward once more during the $\Lambda$-dominated era, forming a second peak.}
    \label{fig:omegaGWsketch}
\end{figure}

More recent work, Ref.~\cite{Figueroa:2019paj}, has shown that instead of kination, a stiff period with a constant barotropic parameter $\frac{1}{3}<w<1$ can produce a gentler gradient than kination proper, namely \cite{Gouttenoire:2021jhk}
\begin{equation}
    \Omega_{GW} \propto f^{-2\left(\frac{1-3w}{1+3w}\right)}\,.
    \label{spectral}
\end{equation}
This is expected to relax the restrictions on the length of the period. 
However, to maintain a constant $w$, a delicate balance between the kinetic and potential energy densities is required, as  evident from Eq.~\eqref{wearly}
In most cases, this seems artificial (see, however, Ref.~\cite{Dimopoulos:2022mce}). This is why, in this work, we aim to demonstrate that a stiff period of gradually varying barotropic parameter (from $-1$ to 1), caused by a gradually thawing scalar field on a steeply inclined potential, has a similar effect on this constraint, producing, instead of a rising slope, a distinctive rounded peak at the high end of the GW energy density spectrum.\footnote{An additional bound on the duration of kination (or another stiff era) is due to the danger of the freefalling scalar field condensate becoming overwhelmed by its own perturbations. As the Universe expands, the perturbations are diluted less drastically than the condensate \cite{Eroncel:2025bcb}. Indeed, the perturbations behave identically to the GW amplitude, so a gradual increase of the barotropic parameter from -1 to 1 equally ameliorates this constraint.}  Such a unique observational signature could be observed in the near future. Fig.~\ref{fig:omegaGWsketch} shows sketches of the different spectra, demonstrating the merits of such a setup.

In our paper, we build a concrete example model that demonstrates the viability of the above scenario. We consider hybrid inflation, where evolution first proceeds in the direction of a primordial inflaton field, but later turns into the direction of a waterfall field, a runaway flat direction in field space. To maximize the amplitude of the stochastic gravitational wave background, we assume inflation at the scale of grand unification (GUT-scale), but we remain agnostic to its details.
We take the waterfall field's potential to have the shape of a double exponential, which can be motivated by string theory. We show that the double exponential results in the desired gradual increase in the barotropic parameter, which in turn results in a characteristic rounded peak in the GW spectrum. Furthermore, we find that at late times, the field adopts a pseudo-scaling behaviour on the runaway potential, which allows its energy density to slowly decay away in a natural manner.

Our paper is structured as follows. In Sec.~\ref{sec:model} we introduce the model and its expected behaviour and give an overview of our numerical treatment. In Sec.~\ref{sec:results} we show the results of our numerical analysis and the form of the primordial gravitational wave spectrum. Finally, in Sec.~\ref{sec:conclusions} we discuss our results and conclusions.

Throughout the paper we employ natural units with 
\mbox{$c=\hbar=k_B=1$} and Newton's gravitational constant \mbox{$8\pi G=\m^{-2}$}, with \mbox{$\m=2.43\times 10^{18}\,$GeV} being the reduced Planck mass.

\section{The Model}
\label{sec:model}

\subsection{Field Content and Primordial Inflation}
\label{subsec:lagrangianfieldeq}

The Lagrangian density of our model is
\begin{equation}
\mathcal L=-\frac12\partial_\mu\sigma\partial^\mu\sigma-\frac12\partial_\mu\phi\partial^\mu\phi-
    W(\sigma,\phi)\,,
\label{L}    
    \end{equation}
with
\begin{equation}
W(\sigma,\phi)=U(\sigma)+V(\phi)+\Delta V(\sigma,\phi)\,,
\end{equation}
where $U(\sigma)$ and $V(\phi)$ are the potentials of the inflaton field $\sigma$ and the waterfall field $\phi$ and $\Delta V(\sigma,\phi)$ is their interaction term. At the background level, the metric takes the Friedmann-Lema\^{i}tre-Robertson-Walker (FLRW) form
\begin{equation} \label{eq:FLRW_metric}
    \dd s^2 = -\dd t^2 + a^2(t) \dd x^2 \, ,
\end{equation}
where $t$ is the cosmic time and $a$ is the scale factor; the equations of motion become
\begin{gather}
    \label{eq:sigma_eom}
    \ddot{\sigma} + 3H\dot{\sigma} + \partial_\sigma U(\sigma) + \partial_\sigma \Delta V(\sigma,\phi) = 0 \, , \\
    \label{eq:phi_eom}
    \ddot{\phi} + 3H\dot{\phi} + \partial_\phi V(\phi) + \partial_\phi \Delta V(\sigma, \phi) = 0 \, , \\
    \label{eq:H}
    3H^2\m^2 = \frac{1}{2}\dot{\sigma}^2 + \frac{1}{2}\dot{\phi}^2 + W(\sigma,\phi) \, ,
\end{gather}
where dots denote derivatives with respect to $t$ and $H=\dot{a}/a$ is the Hubble parameter.

We assume the inflaton potential to feature a gentle slope, e.g., characterised by a Coleman--Weinberg form $U(\sigma) \propto \ln(\frac{\sigma^{2}}{Q^{2}}+1)$, where $Q$ is some renormalisation scale, as in supersymmetric hybrid inflation \cite{Dvali:1994ms}. This slope results in the usual slow-roll inflation, and we assume it conforms to the CMB observations at large scales.

\begin{figure}
\centering
\vspace{-8cm}
\includegraphics[width=1.0\textwidth]{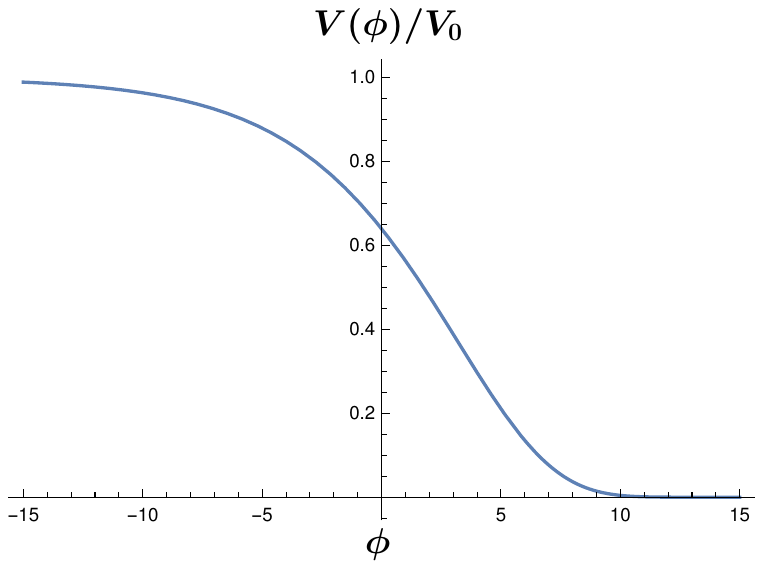}
\vspace{-6.5cm}
    \caption{Schematic representation of the $V(\phi)$ potential given in Eq.~\eqref{V}. The units appearing in the graph are fiducial.}
    \label{fig:expexp2Dplot}
\end{figure}

The waterfall field $\phi$ is a spectator during primordial inflation\footnote{Throughout the paper, `primordial inflation' refers to inflation driven by $\sigma$; it may be followed by a short period of additional inflation driven by the waterfall field $\phi$.}. We take its potential to be 
\begin{equation}
V(\phi) = V_{0}  \;\exp
\left(-\lambda e^{\kappa\,\phi/m_{P}}\right),
\label{V}
\end{equation}
where $\kappa$ and $\lambda$
are dimensionless constants and $V_0$ is a constant energy density scale. The shape of this potential is shown in Fig.~\ref{fig:expexp2Dplot}.
This type of potential is naturally obtained in many string theory realisations \cite{Cicoli:2023opf}.
The second exponential in Eq.~\eqref{V} can be approximated by a Taylor series in the $\kappa\phi<m_{P}$ region,
\begin{align}
    V
    & \simeq V_{0}\exp{-\left[\lambda + \lambda\kappa\frac{\phi}{m_{P}}+\frac12\,\lambda\kappa^2\left(\frac{\phi}{m_{P}}\right)^{2}+\cdots\right]}\, .
\end{align}
The potential decays exponentially, with a rate of decay that grows exponentially. 

The interaction term is
\begin{equation}
    \Delta V(\sigma,\phi) = \frac12g^{2}\phi^2\sigma^2 \, ,
    \label{DV}
\end{equation} 
where $0 < g <{\cal O}(1)$ for perturbativity.
The coupling constant $g$ is assumed to be large enough to pass most of the inflaton's energy density to the waterfall field at the end of primordial inflation
\cite{Garcia-Bellido:1997hex}.

\begin{figure}
\centering
\vspace{-9cm}
\includegraphics[width=1.1\textwidth]{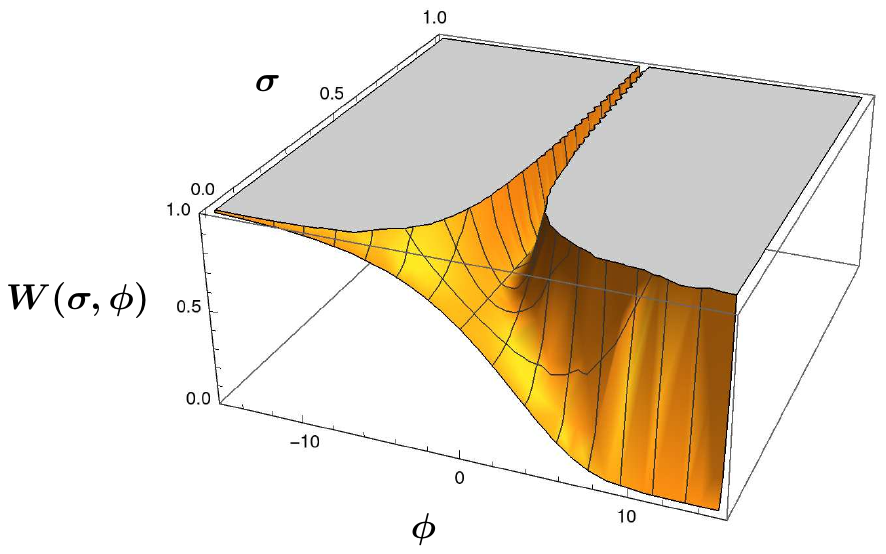}
    \vspace{-7cm}
    \caption{Schematic representation of the $W(\sigma,\phi)$ potential. For $\sigma>0$, the system is driven inside a deep valley with $\phi=0$ as dictated by the potential $\Delta V$. The valley is characterised by a gentle slope given by $V(\sigma)$, which brings the system towards $\sigma=0$. There, the valley opens up and the system slides along the $\phi$-direction, following the potential $V(\phi)$ given in Eq.~\eqref{V}. The units appearing in the graph are fiducial.}
    \label{fig:expexpPlot}
\end{figure}

The full potential with the interaction term is depicted in Fig.~\ref{fig:expexpPlot}. During primordial inflation, the inflaton $\sigma$ is large and the interaction potential \eqref{DV} keeps $\phi$ close to zero. At the end of this period, $\sigma$ approaches zero and starts to oscillate. The expectation value of the waterfall field $\phi$, no longer bound by the interaction, starts to grow in turn, and the interaction quickly pins the inflaton down to the origin. After this phase transition, $\phi$ is oblivious to the interaction and simply rolls down its potential, given in Eq.~\eqref{V}, starting near the origin. Since the inflaton plays no role at this stage, we omit it in our numerical analysis below and concentrate on the rolling waterfall field.

\subsection{After Primordial Inflation}
\label{subsubsec:waterfall}

After the phase transition releases the waterfall field from the origin, the field comes to dominate the Universe. At some point, however, the Universe has to reheat and transition to standard cosmology dominated by radiation and matter. In the original hybrid scenario in Ref.~\cite{Linde:1993cn}, oscillations of the waterfall field brought about reheating. In our scenario, the waterfall field does not oscillate but instead rolls down its runaway direction, similar to Ref.~\cite{Dimopoulos:2022mce}, and the Universe has to reheat by other means.\footnote{Some radiation can be produced by the decay of the oscillating inflaton field $\sigma$, but such radiation is diluted drastically by a limited period of inflation driven by the waterfall field $\phi$. The efficiency of such a mechanism is also limited by the fact that most of the energy density of primordial inflation is passed on to the waterfall field, away from the inflaton.} Several mechanisms have been proposed to this end. Probably the least complicated of these proceeds through an additional spectator field as curvaton reheating \cite{Feng:2002nb,BuenoSanchez:2007jxm} or, if the spectator is non-minimally coupled to gravity, as Ricci reheating \cite{Dimopoulos:2018wfg,Opferkuch:2019zbd,Bettoni:2021zhq}. Other possibilities also exist, for example, reheating due to the evaporation of primordial black holes \cite{Dalianis:2021dbs}.\footnote{Such primordial black holes might be generated at the end of hybrid inflation \cite{Garcia-Bellido:1997hex} but their abundance is expected to be small in our scenario, because the waterfall field is not initially on the top of a potential hill.} In this work, we are agnostic to the details, and simply assume that some mechanism generates the thermal bath of the hot big bang, which eventually comes to dominate and reheats the Universe.

After the thermal bath has been generated, the Universe consists of radiation, cold matter (baryonic and dark), dark energy, and the waterfall field, whose energy densities we denote as $\rho_r$, $\rho_m$, $\rho_\Lambda$, and $\rho_\phi$, respectively. We also introduce the combined `background' energy density $\rho_b=\rho_r+\rho_m+\rho_\Lambda$ for convenience. We denote the Universe's total energy density simply as $\rho$. The pressures $p_x$ and the barotropic parameters $w_x\equiv p_x/\rho_x$ follow the same subscript convention; we have $w_r=1/3$, $w_m=1$, and $w_\Lambda=-1$. For $\phi$, we have
\begin{equation}
    \rho_\phi = K(\phi) + V(\phi) \, ,
    \qquad
    p_\phi = K(\phi) - V(\phi) \, ,
    \qquad
    w_\phi=\frac{K(\phi)-V(\phi)}{K(\phi)+V(\phi)}\,,
    \label{w}
\end{equation}
where \mbox{$K(\phi)\equiv\frac12\dot\phi^2$} is the field's kinetic energy density.

The combined equations of motion for the waterfall field and the matter components read
\begin{gather}
    \label{eq:phi_eom_2}
    \ddot{\phi} + 3H\dot{\phi} + \partial_\phi V(\phi) = 0 \, , \\
    \label{eq:fluid_eom}
    \dot{\rho}_x + 3H\rho_x(1+w_x) = 0 \, , \\
    \label{eq:H_b}
    3H^2\m^2 = \frac{1}{2}\dot{\phi}^2 + V(\phi) + \rho_b \, ,
\end{gather}
where $x=r,m,\Lambda$. Initially, radiation dominates over the other background components.

Let us discuss the various stages of evolution in more detail. They are characterized by the total barotropic parameter $w=p/\rho$.

\begin{enumerate}
\addtocounter{enumi}{-1}

\item{\boldmath\textbf{Inflation from the waterfall field ($-1<w<-\frac13$)}}

Because the waterfall field begins with a very small initial velocity (determined solely by random quantum fluctuations), and because the potential in Eq.~\eqref{V} becomes gradually steeper, there may be a very small number of efolds where its barotropic parameter \mbox{$w_\phi\approx w$} is smaller than $-\frac13$. In this fashion, inflation may continue for a short time along the waterfall direction.\footnote{These extra efolds of inflation may help the model have better agreement with observations, since hybrid inflation with a Coleman--Weinberg potential gives rise to a spectral index of scalar curvature perturbations that is not red enough, \mbox{$n_s\simeq 0.98$}. A brief period of subsequent inflation along the waterfall direction is expected to bring this number down somewhat \cite{Dimopoulos:2016tzn}.}
The contribution to the energy density budget from the inflaton $\sigma$, which oscillates around the origin, soon becomes 
suppressed compared with that of the waterfall field $\phi$. As time passes, the $\phi$-field rolls faster and faster, with an increasing barotropic parameter~$w_\phi$.

\item{\boldmath\textbf{Intermediate period ($-\frac{1}{3}\leq w\leq \frac{1}{3}$)}}

As $w_\phi$ increases, it soon becomes larger than $-\frac13$ and inflation ends. From then on, lengthscales which exited the horizon during inflation start reentering again.
This period corresponds to the higher-frequency side of a peak in the GW energy density spectrum, where $\Omega_{GW}(f)$ is a decreasing function of the frequency~$f$ (from Eq.~\eqref{spectral}, the slope $\dd \ln \Omega_{GW}/\dd \ln f \in [-\infty,0]$). This region is denoted as `STAGE~1' in Fig.~\ref{fig:rounded} and extends to maximum frequency $f_{\rm max}$, which corresponds to the end of inflation.

\begin{figure}
\centering
\vspace{-10cm}
\mbox{\hspace{-1cm}\includegraphics[width=1.2\textwidth]{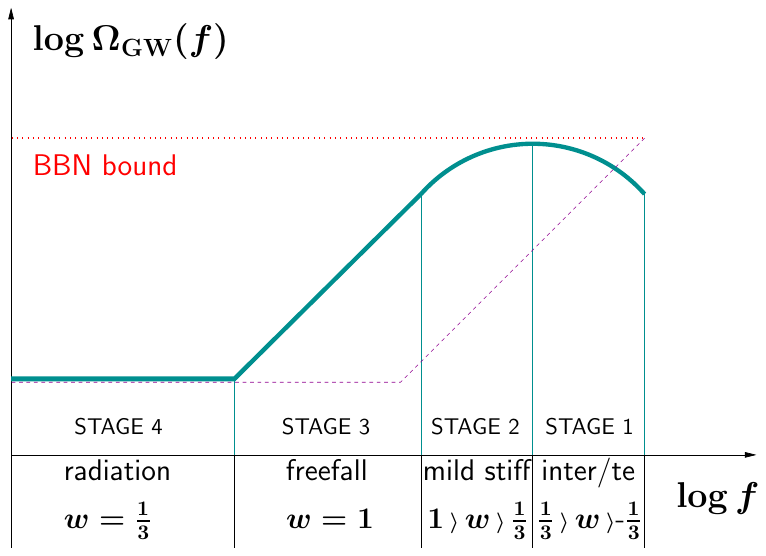}}
\vspace{-8cm}
    \caption{Schematic graph of the expected form of the spectral density parameter $\Omega_{GW}(f)$ as a function of frequency $f$ when the barotropic parameter of the Universe $w(t)$ is growing gradually from $-\frac13$ up to unity, and then falls to $\frac13$ in the radiation era. Stages 1-4 are explained in the text. The dashed line depicts the peak in GWs when kination proper, with $w=1$, immediately follows the end of inflation. In both cases, the GW spectra saturate the $\Delta N_{\textnormal{eff}}$ constraint, depicted by the horizontal dotted line. We see that the case of a rounded peak, corresponding to a gradually increasing $w(t)$, leads to an enhancement of lower frequencies, which facilitate better contact with forthcoming GW observations.
 }
        \label{fig:rounded}
\end{figure}

\item{\boldmath\textbf{Mild stiff period ($\frac{1}{3}<w<1$)}}

The mild stiff period of the field corresponds to an epoch when the density of the field is redshifting faster than radiation but its varying barotropic parameter has not yet hit unity. This period is responsible for the gradual curve at the end of the energy density spectrum for gravitational wave modes. In contrast to the previous period, $\Omega_{GW}(f)$ is increasing with $f$ ($\dd \ln \Omega_{GW}/\dd \ln f \in \ ]0,1]$).
Ideally, we would like this period to be as long as possible within observational bounds, as this would produce a wider and larger peak which 
would be accessible to upcoming observations. This region is denoted as `STAGE~2' in Fig.~\ref{fig:rounded}. 

\item{\boldmath\textbf{Freefall ($w=1$)}}

The waterfall field undergoes a period of freefall, where its energy density decays away as $\rho_{\phi}\propto a^{-6}$. During the period, the field 
is dominated solely by its kinetic energy density $K(\phi)$, and the evolution becomes oblivious to the potential $V(\phi)$. This is why the field is said to undergo freefall motion in field space. Since radiation scales as $\rho_r \propto a^{-4}$, it eventually catches up with the field: reheating occurs and radiation becomes the dominant component of the~Universe. For a little while, the waterfall field continues to freefall, with \mbox{$w_\phi=1>w \approx w_r=\frac13$}. This region, up to reheating, is denoted as `STAGE~3' in Fig.~\ref{fig:rounded}. Up to reheating, the GW spectrum is still increasing with $\dd \ln \Omega_{GW}/\dd \ln f =1$; after reheating, the spectum is flat, $\dd \ln \Omega_{GW}/\dd \ln f =0$.

\item{\boldmath\textbf{Freezing ($w_\phi=-1$ \& $w=\frac13$)}}

The kinetic energy density of the waterfall field is rapidly depleted by excessive Hubble friction, which, after reheating, is determined by the less rapidly diluting radiation background. As a result, we soon have \mbox{$K(\phi)\ll V(\phi)$} and the field freezes at a value $\phi_F$ with the energy density \mbox{$\rho_\phi\simeq V(\phi_F)=\,$constant} and the barotropic parameter \mbox{$w_\phi\simeq -1$}, provided \mbox{$\partial_{\phi}^{2}V(\phi_F)\ll H^2$}. This region, the radiation era of the hot big bang, is denoted as `STAGE~4' in Fig.~\ref{fig:rounded} and corresponds to a flat GW spectrum.

\item{\boldmath\textbf{Thawing and pseudo-scaling solution ($w_\phi\approx w=\frac13$ and beyond)}}

At late times, the decreasing Hubble parameter $H(t)$ becomes comparable to the field's effective mass, \mbox{$\partial_{\phi}^{2}V(\phi_F)\sim H^2$}, and we expect the waterfall field to
unfreeze. Afterwards, it turns out that $\phi$
scales with the dominant component of the Universe's matter content, which is first radiation, but later switches to matter and dark energy, in an attractor-like manner. However, this is not a true attractor, but instead a ``pseudo-attractor,'' because the attractor behaviour is asymptotic. This may be understood as follows.

Let us introduce the tracking parameter,
defined as \cite{Copeland:2006wr}
\begin{equation}
    \Gamma \equiv \frac{V(\phi)\partial_{\phi}^{2}V(\phi)}{[\partial_{\phi}V(\phi)]^{2}} \, .
\end{equation}
A scalar field eventually adopts attractor behaviour if the tracking parameter satisfies $\Gamma \geq 1$. On the attractor, the field's barotropic parameter and energy density fraction become \cite{Copeland:2006wr}
\begin{equation} \label{eq:attractor_w_Omega}
    w_{\phi} = \frac{w_b - 2(\Gamma-1)}{2\Gamma-1} \, , \qquad
    \Omega_{\phi} \equiv \frac{\rho_\phi}{\rho_\phi+\rho_B} = \frac{3}{\ell^{2}}(w_{\phi}+1) \, ,
\end{equation}
where \mbox{$
\ell\equiv-\frac{m_{P}\partial_{\phi}V(\phi)}{V(\phi)}$} is the strength of the potential described in terms of an exponential. If $\Gamma = 1$, this is a scaling attractor with $w_\phi=w_b$. The field energy density can be either dominant or subdominant compared to the background energy density \cite{Copeland:1997et}, following Eq.~\eqref{eq:attractor_w_Omega}.

Even if $\Gamma < 1$ everywhere in a strict sense, we postulate that a pseudo-scaling attractor exists if 
\begin{equation}
    \lim_{\phi\rightarrow\infty}\Gamma = 1,
\end{equation}
so that in the large-field limit, the potential asymptotes to one supporting a scaling attractor (i.e. an exponential \cite{Copeland:1997et}). As a result, the field tends towards, but never quite reaches, the scaling behaviour: $w_\phi$ approaches $w_b$ and $\Omega_\phi$ approaches a constant, but at a rate that may be slow. To the best of our knowledge, such behaviour has not been studied previously in the literature.

For our potential,
\begin{equation}
    \Gamma = 1- \frac{1}{\lambda e^{\kappa\phi/m_{P}}} \xrightarrow{\phi \to \infty} 1 \, ,
\end{equation}
so we expect pseudo-attractor behaviour at late times. Figs.~\ref{fig:barotropic} and~\ref{fig:densityparameter} show this in an example case. At late times, $w_\phi$ approaches the background barotropic parameter but stays above it. The density parameter $\Omega_\phi$ is subdominant and keeps decreasing slowly during radiation domination, but plateaus during matter and dark energy domination where the field is presumably almost on an attractor. The field seems to follow the changes in the attractor very well, based on the dominant component of the Universe. We leave further study of this behaviour to future work. 
\end{enumerate}

\subsection{Solving Homogeneous Dynamics}
\label{subsec:numericalsimulation}
To verify the behaviour described in Sec.~\ref{subsubsec:waterfall}, we wish to solve the system of equations \eqref{eq:phi_eom_2}--\eqref{eq:H_b} numerically. We follow the system from inflation until today with Mathematica (we identify the time of today in the simulations by the correct density parameter of the cosmological constant $\Omega_{\Lambda} = 0.6889$ from Planck 2018 \cite{Planck:2018vyg}). The input parameter values and initial conditions are collected in Table~\ref{tab:backgroundinitcond}.

\begin{table}
\centering
    \begin{tabular}{|p{3.5cm}|p{4.5cm}|p{4.5cm}|}
    \hline
        \textbf{Parameter} & \textbf{Initial Value} & \textbf{Explanation}\\
        \hline
        \footnotesize{Model parameters $\lambda$, $\kappa$} & \footnotesize{Varied in parameter space scans} & \\ \hline
        \footnotesize{Model parameter $V_{0}$} & \footnotesize{$V_0=V_{\textnormal{ini}}=1.9 \times10^{-9} m_{P}^{4}$} & \footnotesize{Maximal inflationary scale allowed by \cite{Planck:2018jri} (GUT inflationary scale), maximizes SGWB amplitude} \\ \hline
       \footnotesize{Initial field value} & $\phi_{\textnormal{ini}}  = \frac{H_{\textnormal{ini}}}{2\pi}$ & \footnotesize{From quantum fluctuations at waterfall transition}\\
        \hline
        \footnotesize{Initial field velocity} & $\phi'_{\textnormal{ini}} = \frac{H_{\textnormal{ini}}}{2\pi}$ & \footnotesize{From quantum fluctuations at waterfall transition
        }
        \\
        \hline
        \footnotesize{Initial radiation energy density $\rho_{r}(0)$} &  \footnotesize{Varied as a model parameter} & \footnotesize{Sets the moment of reheating
        }\\
        \hline
        \footnotesize{Initial matter energy density} & \footnotesize{\mbox{$\rho_{m}^{\textnormal{ini}} = \rho_{m}^{\textnormal{eq}}\exp\left[\frac{3}{4}\ln(\frac{\rho_{r}^{\textnormal{ini}}}{\rho_{m}^{\textnormal{eq}}})\right]$} where $\rho_{m}^{\textnormal{eq}} = 3\Omega_{m}^{0}H_{0}^{2}(z_{\textnormal{eq}}+1)^{3}$ as measured by Planck 2018 (i.e. $\Omega_{m}^{0}=0.3111$, $H_{0}=67.66 \ \textnormal{km/s} \ \textnormal{Mpc}$, $z_{\textnormal{eq}} = 3387$) \cite{Planck:2018vyg}} & \footnotesize{Chosen such that matter--radiation equality occurs at the correct energy density ($10^{-108} m_{P}^{4}$)}\\
        \hline
        \footnotesize{Cosmological constant energy density} & \footnotesize{$\rho_{\Lambda}= 3H_{0}^{2}\Omega_{\Lambda}$ as measured by Planck 2018 \cite{Planck:2018vyg} (where $\Omega_{\Lambda} = 0.6889$ and $H_{0}=67.66 \ \textnormal{km/s} \ \textnormal{Mpc}$)} & \footnotesize{Set to be consistent with the Planck 2018 \cite{Planck:2018vyg} measurement of dark energy}\\
        \hline
    \end{tabular}
    \caption{Initial conditions and parameter values for the background dynamics.}
    \label{tab:backgroundinitcond}
\end{table}

Since the evolution spans a wide range of time and energy scales, we take steps to improve the system's numerical behaviour.

\paragraph{\boldmath{$N$} as the time variable.} Instead of the cosmic time $t$, we use the number of efolds $N=\ln a(t)$ as the time variable. Writing $\dd N = H \dd t$, we get $\dot{\rho}_x = H\rho_x'$, where prime denotes a derivative with respect to $N$. This lets us integrate the background continuity equations \eqref{eq:fluid_eom}:
\begin{equation} \label{eq:rho_x_scaling}
    \rho'_x = -3(1+w_x)\rho \quad \implies \quad
    \rho_x = \rho_x(0) e^{-3N(1+w_x)} \, , \qquad x=r,m,\Lambda \, ,
\end{equation}
where $\rho_x(0)$ are set by initial conditions.

For the waterfall field, we obtain $\dot{\phi}=H\phi'$ and $\ddot{\phi} = H^2\phi'' + \dot{H}\phi'$. The field equation \eqref{eq:phi_eom_2} becomes
\begin{equation} \label{eq:phi_N_pre_eom}
    \phi'' + \left(\frac{\dot{H}}{H^{2}} + 3\right)\phi' + \frac{\partial_{\phi}V}{H^{2}} = 0 \, .
\end{equation}
We deal with the Hubble factors $H$ and $\dot{H}$ with the Friedmann equations.

The first Friedmann equation \eqref{eq:H_b} becomes
\begin{equation}
    \label{eq:H_2}
    3H^2\m^2 = \rho = \rho_\phi + \rho_b
    =\frac{1}{2}H^2\phi'^2 + V(\phi) + \rho_b
\end{equation}
(remember $\rho_b$ contains all the background components), and using this and the definition of $p_\phi$ from Eq.~\eqref{w} lets us solve
\begin{equation}
    \label{eq:rho_p_in_N}
    \rho_{\phi} = \frac{6\m^{2}V + \rho_{b}(\phi')^{2}}{6\m^{2}-(\phi')^{2}} \,,
    \qquad
    p_{\phi} = \frac{1}{2}\left(\frac{\rho_{\phi}+\rho_{b}}{3m_{P}^{2}}\right)(\phi')^{2} - V(\phi) \, .
\end{equation}
These expressions are useful for keeping track of the waterfall field's energy density and equation of state; additionally, plugging Eq.~\eqref{eq:rho_p_in_N} back into Eq.~\eqref{eq:H_2} gives us $H^2$ in terms of the dynamic variables $\phi$, $\phi'$, and $\rho_b$.

For $\dot{H}$, we use the second Friedmann equation $\ddot{a}/a = -(\rho + 3p)/(6\m^2)$ together with the first to obtain
\begin{equation}
\label{eqn:friedmann2}
    \dot{H} = \frac{\ddot{a}}{a} - H^2
    = -\frac{1}{2\m^2}(\rho + p) = -\frac{1}{2\m^2}(\rho_\phi + \rho_b + p_\phi + p_b) \, .
\end{equation}
All in all, Eq.~\eqref{eq:phi_N_pre_eom} becomes
\begin{equation} \label{eq:phi_N_eom}
    \phi'' + \frac{3}{2}\left(1-\frac{p_\phi + p_b}{\rho_\phi + \rho_b}\right)\phi' + \frac{dV}{d\phi}\frac{3m_{P}^{2}}{\rho_{\phi}+\rho_b} = 0 \, .
\end{equation}
This is a closed equation for $\phi$ once the energy densities and pressures are plugged in using Eqs.~\eqref{eq:rho_x_scaling} and \eqref{eq:rho_p_in_N}.

\paragraph{Logarithmic energy scale.} 
To avoid extremely small or large numbers in the numerics, we treat the energy densities logarithmically. In particular, in the denominators of Eq.~\eqref{eq:phi_N_eom}, we write
\begin{equation}
    \rho_\phi + \rho_b = \rho_b\qty[\exp(\ln{\rho_\phi} - \ln{\rho_b} ) + 1] \, ,
\end{equation}
which turns out to improve the behaviour of the computation. This gives the final form of our numerical equations.

\subsection{Solving Gravitational Wave Spectrum}
\label{subsec:mukhanovsasaki}

In Sec.~\ref{subsubsec:waterfall}, we discussed our expectations for the SGWB spectrum in our model. To confirm these expectations, we solve the gravitational wave mode equation numerically, using the numerical solution of background dynamics.

\paragraph{Mode equation.} We start by studying the transverse-traceless tensor perturbation $h_{ij}$. At linear order, its time evolution is governed by the sourceless Mukhanov--Sasaki equation
\begin{equation}
    \partial^2_\tau{h}_{ij} + 2aH\partial_\tau{h}_{ij} - \nabla^2 h_{ij} = 0 \, ,
    \label{eq:sourcelessMukhanov}
\end{equation}
where the new time variable $\tau$ is the conformal time, defined through $\dd t = a\dd \tau$ \cite{Boyle:2005se}. We can Fourier expand the tensor perturbations and decompose them into two polarization modes as\footnote{This equation is schematic: in a more detailed treatment, $h_{ij}$ is a quantum field and $h_k^s$ are mode functions multiplying the ladder operators. We omit the details which are standard \cite{Lyth:2009imm}.}
\begin{equation}
    h_{ij} (\tau, {\bf x}) =\int \frac{d^3k}{(2\pi)^{3/2}} \sum_{s=+,\times} h^s_k(\tau)
    e^s_{ij} ({\bf k}) e^{i{\bf k\cdot x}} \, ,
\end{equation}
where $e^s_{ij}$ is symmetric (\mbox{$e^s_{ij}=e^s_{ji}$}), traceless (\mbox{$e^s_{ii}=0$}), and transverse (\mbox{$k^i e^s_{ij}=0$}).
From now on, we suppress the polarization index `$s$'. The time evolution is contained in the mode amplitudes $h_k$, which follow
\begin{equation}
\partial^2_\tau{h}_k + 2aH\partial_\tau{h}_k + k^{2}h_k = 0\,.
    \label{eq:hkeqn}
\end{equation}
We again wish to solve this equation numerically, in terms of the efolds $N$. The change of variables $\dd N = aH\dd \tau$ gives
\begin{equation}
    \partial_\tau h_{k} = aHh'_{k}\,,
    \quad
    \partial^2_\tau h_{k} = \left(a^{2}H\frac{dH}{dN} + a^{2}H^{2}\right)h_{k}' + a^2H^2 h_{k}''\, ,
\end{equation}
and using also the first Friedmann equation and $a=e^N$, Eq.~\eqref{eq:hkeqn} becomes
\begin{equation} \label{eq:h_eom}
    \frac{\rho}{3}h_{k}'' + \left(\frac{1}{6}\frac{d\rho}{dN} + \rho\right)h_{k}' + \m^2k^{2}e^{-2N}h_{k} = 0 \, ,
\end{equation}
which we can integrate after solving the background behaviour $\rho(N)$.

\paragraph{Solving the mode equation.}
During cosmic inflation, assuming the Hubble parameter $H=H_\text{inf}$ is approximately constant, the mode functions follow the solution
\begin{equation}
h_k=\frac{1}{a\sqrt{2k}}
\left(1-\frac{i}{k\tau}\right)
e^{-ik\tau}. 
\label{eqn:solu}
\end{equation}
This is chosen to comply with the Bunch--Davies initial conditions in the sub-Hubble limit $k\tau \to -\infty$ \cite{Boyle:2005se}. In the opposite limit, well after Hubble exit, the mode function freezes to
\begin{equation} \label{eq:mode_frozen}
     h_{k} \xrightarrow{k\tau \to 0} -\frac{i}{a\sqrt{2k}}\frac{1}{k\tau} = 
     \frac{i}{\sqrt{2k}}\frac{H_{\textnormal{inf}}}{k} \, , \qquad
     h'_k \xrightarrow{k\tau \to 0} 0 \, ,
\end{equation}
where we used $\tau = -1/(aH_\text{inf})$. We take $H_{\textnormal{inf}}=\sqrt{\frac{V_\text{ini}}{3\m^{2}}}$, where $V_{\textnormal{ini}}=1.9 \times10^{-9} m_{P}^{4}$ is the initial value of the waterfall field potential.

We start our numerical simulations from the frozen state in Eq.~\eqref{eq:mode_frozen} at super-Hubble scales and evolve the modes with Eq.~\eqref{eq:h_eom} up to three efolds beyond the mode's horizon entry. We repeat this for a wide range of $k$-values spaced logarithmically.

\paragraph{Energy density of gravitational waves.}
\label{subsubsec:GWenergydensity}

We use the numerically obtained mode functions to compute the energy density spectrum of the gravitational waves. The spectral density parameter is \cite{Boyle:2005se}
\begin{align}
    \Omega_{GW}(k,N) &= \frac{1}{\rho(N)}\frac{k^{3}}{2\pi^{2}}\frac{|\partial_\tau h_k(N)|^{2} + k^{2}|h_{k}(N)|^{2}}{a^{2}(N)} \\
    &= \frac{k^{3}}{2\pi^{2}}\left(\frac{1}{3\m^2}\left|\frac{dh_{k}}{dN}\right|^{2} + \frac{k^{2}}{\rho(N)a^{2}(N)}|h_{k}|^{2}\right) \, ,
\end{align}
and it gives the energy in a logarithmic bin around $k$. The total density parameter of SGWB is $\Omega_{GW}(N) = \int \dd \ln k \,\Omega_{GW}(k,N)$. We first compute $\Omega_{GW}(k,N)$ at the end time of our numerical mode evolution, that is, three efolds after the mode's horizon entry; afterwards, we utilize the fact that the GW energy density scales as $a^{-4}$ inside the horizon to project $\Omega_{GW}(k,N)$ to present day.

When modelling the background dynamics, we assumed a fixed number of degrees of freedom in radiation, so that $w_r$ stays constant. In reality, phase transitions change the degrees of freedom and leave a mark on the SGWB spectrum. This can be taken into account with a step-function type correction, as established in Ref.~\cite{Figueroa:2019paj}. We make the replacement
\begin{equation}
    \Omega_{GW}(k,N) \to  C_{\Delta g*}(k) \Omega_{GW}(k,N) \, ,
\end{equation}
where the correction factor is
\begin{equation}
    C_{\Delta g*}(k)\equiv \frac{\mathcal{G}_{k}(g_{*,k}, g_{s,k})}{\mathcal{G}_{k}(10.75,10.75)} \, , 
    \qquad
    \mathcal{G}_{k}(g_{*,k}, g_{s,k}) \equiv \left(\frac{g_{*,k}}{g_{s,0}}\right)\left(\frac{g_{s,0}}{g_{s,k}}\right)^{\frac{4}{3}} \, ,
\end{equation}
$g_{*,k}$ is the effective number of relativistic degrees of freedom for relativistic energy density when the scale $k$ re-enters the horizon, $g_{s,k}$ is the same for the entropy of the relativistic component, the subscript `0' refers to the present time, and $10.75$ is our chosen reference value. We model $\mathcal{G}_{k}$ with two steps, one at the QCD phase transition at $T=200$ MeV, before which $g_{*}\simeq g_{s} \simeq 106.75$ and after which $g_{*}\simeq g_{s} \simeq 10.75$, and the other at the electron-positron annihilation at $T = 0.5$ MeV, after which $g_{*} \simeq 3.36$ and $g_{s} \simeq 3.91$.
As noted in Ref.~\cite{Figueroa:2019paj}, this is not a large effect, and produces a correction in all cases around order unity.

\subsection{Observational constraints}
\label{subsec:constraints}
Big Bang Nucleosynthesis is sensitive to deviations from standard cosmology, and hence sets constraints on models. 
After obtaining the solutions for the background evolution and the SGWB spectrum, we check them against $\Delta N_{\textnormal{eff}}$ constraints. We find BBN by the point at which the radiation density has a temperature of $1$ MeV. 

\paragraph{Density parameter of the field at BBN.} The scalar field's high energy density may disrupt BBN \cite{Arbey:2019cpf}. The constraint on the energy density depends on the field's barotropic parameter as follows:\footnote{A more recent study, Ref.~\cite{An:2023buh}, gives the more lenient constraint $\rho_{\phi}/\rho_{r}|_{_{\textnormal{BBN}}}< 0.15$ from $w=1/3$.}
\begin{alignat}{2}
\label{eqn:bbndensityconstraint}
&w_\phi = 1 : \qquad\quad && \frac{\rho_{\phi}}{\rho_{r}}\bigg|_\text{BBN} < 1.40 \, , \\
&w_\phi=-1 : && \frac{\rho_{\phi}}{\rho_{r}}\bigg|_\text{BBN} < 2 \times 10^{-7} \, ,\\
\label{eq:phi_BBN_constraint_rad}
&w_\phi=\frac{1}{3}: && \frac{\rho_{\phi}}{\rho_{r}}\bigg|_\text{BBN} < 0.11 \, .
\end{alignat}
We check each of our simulations for the value of the field's barotropic parameter and apply the relevant constraint if $|w_{\phi}-w_{\textnormal{constraint}}|<0.02$, discarding any other points. In practice, we have found that for the majority of the parameter space tested, the field is following the subdominant attractor at BBN and the barotropic parameter is close to $\frac{1}{3}$.

\paragraph{Integrated density of gravitational waves at BBN.}
The energy density of gravitational waves can also disrupt BBN. In light of this, Ref.~\cite{Maggiore:1999vm} finds the integrated constraint 
\begin{equation}
\int^{k_{\textnormal{end}}}_{k_{\textnormal{BBN}}}\dd \ln{k} \, h_{0}^{2} \, \Omega_{\textnormal{GW}}(k,\text{today}) \leq 5.6 \times 10^{-6} \, ,
\label{eq:gravwaveBBNbound}
\end{equation}
where we use the conservative value $\Delta N_{\textnormal{eff}}=1$ from the corresponding constraint $\Delta N_{\textnormal{eff}} < 1$ for the extra relativistic degrees of freedom (where $k_{\textnormal{end}}$ corresponds to the end of inflation and $k_{\textnormal{BBN}}$ corresponds to BBN). When applying this constraint, we use $h_0=0.6762$ where $h_{0}$ is the dimensionless Hubble constant as found by ACT \cite{ACT:2025fju}. In practice, the constraint constrains the height of the SGWB's high-frequency peak.

\section{Results and Analysis}
\label{sec:results}

\subsection{Field Behaviour}
\label{subsec:fieldbehaviour}

\begin{figure}[H]
    \centering
    \includegraphics[width=\linewidth]{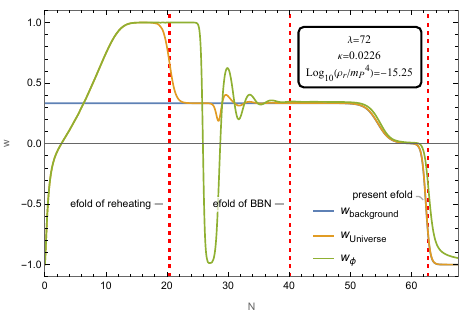}
    \caption{The barotropic parameter of the background (radiation, matter and $\Lambda$) of the Universe (blue), the Universe overall (orange), and the field $\phi$ (green), for the point with model parameters $\lambda = 72$, $\kappa=0.0226$, and $\log_{10}(\rho_{r}^{\textnormal{ini}}/\m^{4})=-15.25$, as a function of elapsing efolds.}
    \label{fig:barotropic}
\end{figure}

\begin{figure}[H]
    \centering
    \includegraphics[width=\linewidth]{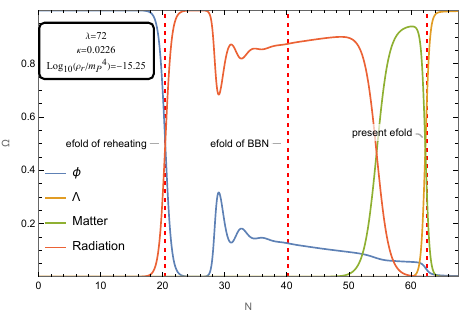}
    \caption{The density parameters of the field $\phi$ (blue), the cosmological constant $\Lambda$ (orange), matter (green) and radiation (red) for the point model parameters $\lambda = 72$, $\kappa=0.0226$, and $\log_{10}(\rho_{r}^{\textnormal{ini}}/\m^{4})=-15.25$, as a function of elapsing efolds. The density parameter of the field marginally saturates the BBN density constraint in Eq. \eqref{eqn:bbndensityconstraint} with $\Omega_{\phi} = 0.1099$.}
    \label{fig:densityparameter}
\end{figure}

\begin{figure}[H]
    \centering
    \includegraphics[width=\linewidth]{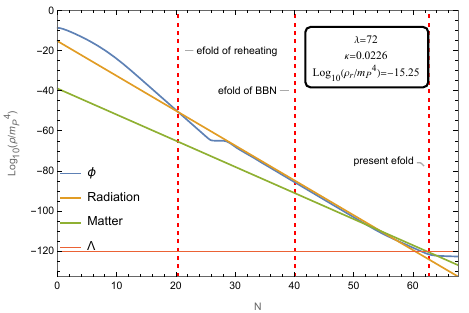}
    \caption{The logarithmic energy densities of the field $\phi$ (blue), radiation (orange), matter (green) and $\Lambda$ (red) for the point with model parameters $\lambda = 72$, $\kappa=0.0226$, and $\log_{10}(\rho_{r}^{\textnormal{ini}}/\m^{4})=-15.25$, as a function of elapsing efolds.}
    \label{fig:energydensities}
\end{figure}

Let us study the evolution of the field for an example point  with model parameters $\lambda = 72$, $\kappa=0.0226$, and $\log_{10}(\rho_{r}^{\textnormal{ini}}/\m^{4})=-15.25$. The evolutions of the barotropic parameters, density parameters, and absolute energy densities are depicted in Figs.~\ref{fig:barotropic}, \ref{fig:densityparameter}, and \ref{fig:energydensities}. The behaviour agrees with the discussion of Sec.~\ref{subsubsec:waterfall}. 

The field starts in inflation after the waterfall transition at $N=0$. The following inflationary period is short, lasting around $\sim 2$ efolds, during which the field thaws from $w=-1$ to $w=-\frac{1}{3}$, see Fig.~\ref{fig:barotropic}. As noted in Sec.~\ref{subsubsec:waterfall}, the small number of extra inflationary efolds can slightly red-tilt the scalar spectral index towards better agreement with CMB observations for Coleman--Weinberg-like inflation.

The subsequent intermediate period with $-\frac{1}{3}\leq w_{\phi} \leq \frac{1}{3}$ 
(STAGE 1 as per the numbering of Sec.~\ref{subsubsec:waterfall} and Fig.~\ref{fig:rounded})
lasts approximately twice as long as the inflationary one, from $N\sim 2$ to $N\sim7$.
Next, the field undergoes an extended stiff period where the barotropic parameter first climbs towards unity
(STAGE 2), which it reaches at $N\sim12$, and then stays there for around 12 efolds (STAGE 3). At $N=20.42$, the radiation energy density catches up with the field and the Universe reheats (STAGE 4), as can be seen in Figs.~\ref{fig:densityparameter}, \ref{fig:energydensities}. The field energy density quickly becomes negligible compared to radiation. The reheating temperature is $\sim 400$ TeV.

Somewhat after reheating, at $N=25.77$, freefall ends and the field briefly freezes, so that $w_\phi=-1$. The field's energy density stays constant, starting to catch up with that of radiation.

The field thaws again as it approaches the pseudo-scaling attractor described in Sec.~\ref{sec:model}. The barotropic parameter $w_\phi$ starts to oscillate around the background's barotropic parameter at $N\sim28$ and settles to it after $\sim 10$ efolds, by $N=40.15$, when BBN happens. The density parameter reaches a peak $\Omega_\phi\sim0.3$ at $N\sim 29$, after which it starts to decrease. At BBN, we have $w_\phi=0.3468$ and $\Omega_\phi=0.1257$ ($\frac{\rho_{\phi}}{\rho_{r}} = 0.1099$), obeying the $\Delta N_{\textnormal{eff}}$ constraint in Eq.~\eqref{eq:gravwaveBBNbound}.

After BBN, the field follows the attractor closely with a decreasing density parameter and a barotropic parameter that always tends towards that of the background, even after the transitions to matter and dark energy dominations at $N=54.48$ and $N\sim62$. The decreasing density parameter ensures that the field has minimal effect on late-Universe dynamics. The present day (identified by $\Omega_{\Lambda} = 0.6889$) corresponds to $N = 62.66$, with $\Omega_\phi=0.0236$.

\subsection{Primordial Gravitational Wave Spectrum}
\label{subsec:tensorperturbationspectrum}

\begin{figure}[H]
   \makebox[\textwidth][c]{
   \hspace{-2.5cm}
   \includegraphics[width=1.3\textwidth]{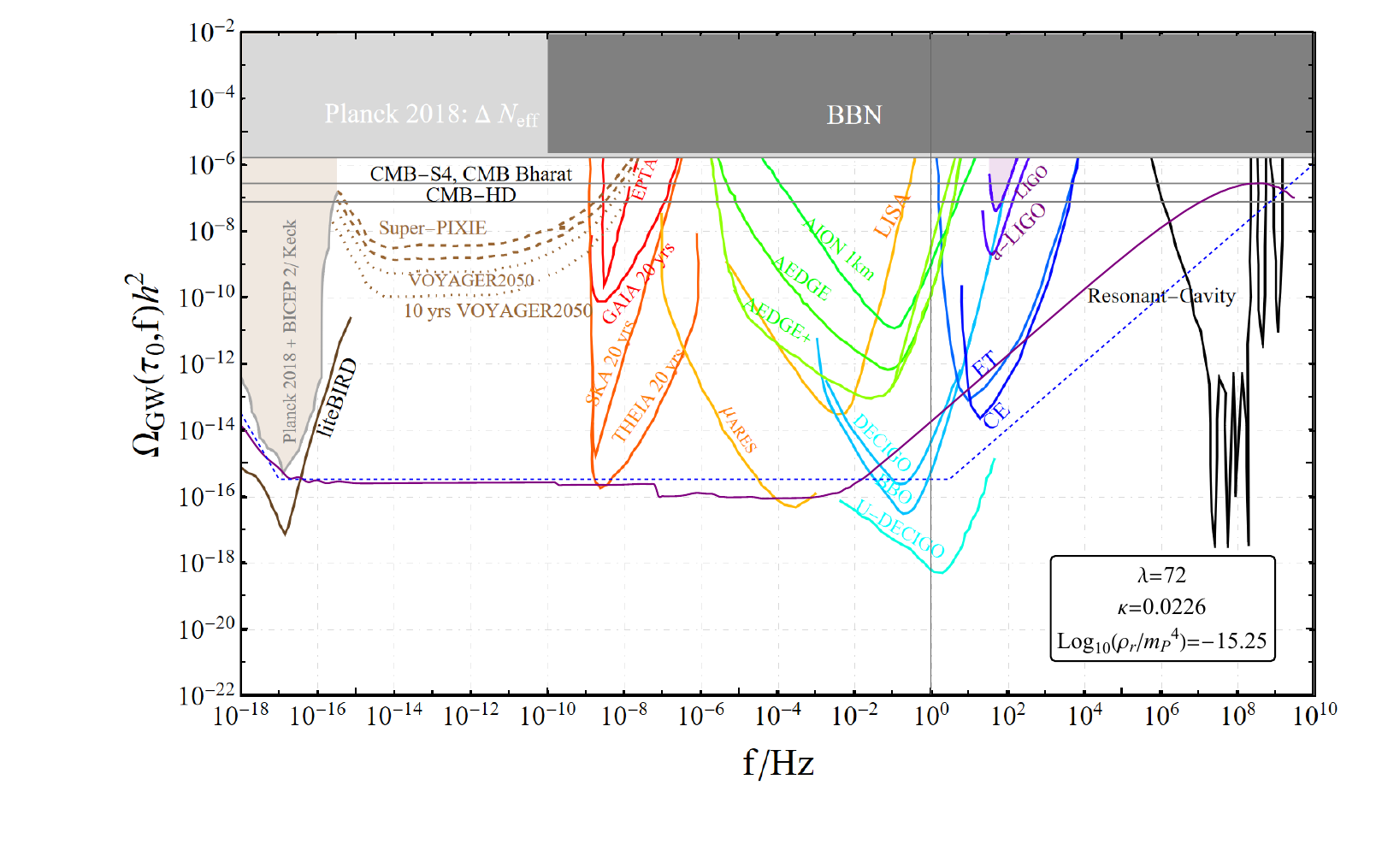}}
    \caption{The solid purple line shows the gravitational wave spectrum of the point with model parameters $\lambda = 72$, $\kappa=0.0226$, and $\log_{10}(\rho_{r}^{\textnormal{ini}}/\m^{4})=-15.25$, superimposed over the sensitivities of upcoming and current observations. For comparison, the dashed blue line is a sketch of the same spectrum with a period of kination which just saturates the $\Delta N_{\textnormal{eff}}$ constraint. 
    The expected sensitivity curves of various
operating and forthcoming GW observatories are also shown, including ground-based interferometer detectors: LIGO/VIRGO \cite{LIGOScientific:2016aoc,LIGOScientific:2016sjg,LIGOScientific:2017bnn,LIGOScientific:2017vox,LIGOScientific:2017ycc,LIGOScientific:2017vwq}, aLIGO/aVIRGO \cite{LIGOScientific:2014pky,VIRGO:2014yos,LIGOScientific:2019lzm}, AION \cite{Badurina:2021rgt,Graham:2016plp,Graham:2017pmn,Badurina:2019hst}, Einstein Telescope (ET) \cite{Punturo:2010zz,Hild:2010id}, Cosmic Explorer (CE) \cite{LIGOScientific:2016wof,Reitze:2019iox};
space-based interferometer detectors: LISA \cite{amaroseoane2017laserinterferometerspaceantenna,Baker:2019nia}, BBO \cite{Crowder:2005nr,Corbin:2005ny}, DECIGO/U-DECIGO \cite{Seto:2001qf,Kudoh:2005as,Yagi:2011wg,Kawamura:2020pcg}, AEDGE \cite{AEDGE:2019nxb}, $\mu$ARES, VOYAGER2050 \cite{Sesana:2019vho}; CMB spectral distortions, PIXIE/Super-PIXIE \cite{Kogut:2019vqh}; recasts of star surveys: GAIA/THEIA \cite{Garcia-Bellido:2021zgu}; CMB polarization measurements, Planck 2018 \cite{Planck:2018jri} and BICEP 2/ Keck \cite{BICEP2:2018kqh,Clarke:2020bil}, LiteBIRD \cite{Litebird}, square-kilometer
array (SKA) \cite{Carilli:2004nx,Janssen:2014dka,Weltman:2018zrl}, EPTA \cite{EPTA:2015qep,EPTA:2015gke}, NANOGRAV \cite{McLaughlin:2013ira,NANOGRAV:2018hou,Aggarwal:2018mgp,Brazier:2019mmu,NANOGrav:2020bcs,NANOGrav:2023hvm,NANOGrav:2023gor}; conversion into electromagnetic waves, the resonant cavity
experiments \cite{Herman:2022fau,Aggarwal:2020olq,Ringwald:2022xif}, and next-generation CMB observations: CMB-S4 \cite{Bianchini:2025dhf}, CMB-Bharat \cite{Adak:2021lbu}, CMB-HD \cite{Sehgal:2019ewc}.
    It is clear that the period of varying barotropic parameter makes better contact with observations, especially with the Einstein Telescope (ET) and the Cosmic Explorer (CE). 
    }
    \label{fig:GWspectrum}
\end{figure}

The gravitational wave energy density spectrum for the point with model parameters $\lambda = 72$, $\kappa=0.0226$, and $\log_{10}(\rho_{r}^{\textnormal{ini}}/\m^{4})=-15.25$ is shown in Fig.~\ref{fig:GWspectrum}. As discussed above, the distinctive feature that this model produces is a rounded peak at the high-frequency end of the spectrum (where modes reentered the horizon earliest), produced by the varying barotropic parameter of the field $\phi$ during its domination. The integrated energy density of gravitational waves is $\int^{f_{\textnormal{end}}}_{f_{\textnormal{BBN}}}d(\log{f})h_{0}^{2}\Omega_{\textnormal{GW}}(f)$ \mbox{$ = 5.0978 \times 10^{-7}$}, which is inside the allowed bound of Eq.~\eqref{eq:gravwaveBBNbound}. In contrast to a period of pure kination, our setup allows the stiff period to continue for longer and brings the GW spectrum in contact with more observations, whilst avoiding violation of the integrated $\Delta N_{\textnormal{eff}}$ constraint. The spectrum of Fig.~\ref{fig:GWspectrum} makes contact with forthcoming observations by the Einstein Telescope and the Cosmic Explorer.

\subsection{Parameter Space}
\label{subsec:paramspace}

\begin{figure}[H]
    \centering
    \includegraphics[width=\linewidth]{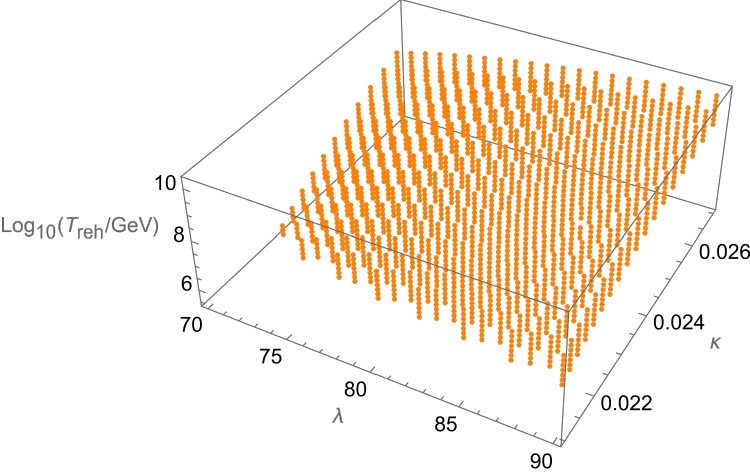}
    \caption{3D parameter space of points which fit the constraints outlined in Sec.~\ref{subsec:constraints}.}
    \label{fig:3dparamspace}
\end{figure}

\begin{figure}
    \centering
    \includegraphics[width=\linewidth]{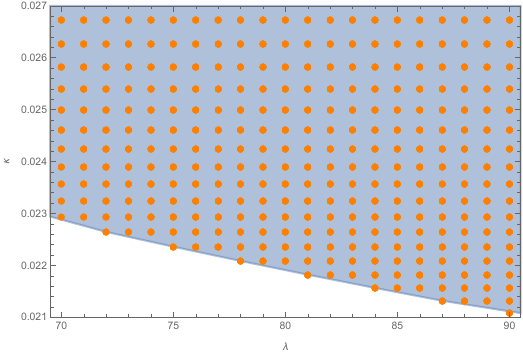}
    \caption{2-D projection of the 3-D parameter space in $\kappa-\lambda$ of points which fit the constraints outlined in Sec.~\ref{subsec:constraints}. We interpolate the underlying parameter space (blue) from the given points (orange).}
    \label{fig:kappalambdaparamspace}
\end{figure}

In Figs.~\ref{fig:3dparamspace} and
\ref{fig:kappalambdaparamspace} we show some of the viable parameter space for the model, in terms of $\lambda$, $\kappa$, and the reheating temperature $T_{\textnormal{reh}}$, subject to the constraints outlined in Sec.~\ref{subsec:constraints}. Fig. \ref{fig:3dparamspace} shows the 3-d parameter space, while Fig. \ref{fig:kappalambdaparamspace} shows the 2-d projection onto the parameters of the scalar field potential.

Along the low-$\lambda$ edge of the parameter space shown, the gravitational wave spectrum is largest. This region of parameter space is most interesting to consider because it is more likely to be observable within the next decade.

Examining the example point considered above, it can be illustrated which constraints are saturated in different directions. In the direction of decreasing $\lambda$, the scalar field has too high an energy density at BBN. In the direction of decreasing reheating temperature, BBN is disturbed by the saturation of the gravitational wave constraint. Finally, in the direction of decreasing $\kappa$, the constraint on the scalar field energy density at BBN is also saturated\footnote{It should be noted that the non-trivial shape of the parameter space at the upper reheating temperature results from the range of reheating temperatures scanned over and is not due to the saturation of any constraint.}.

The parameter space suggests that the values of $\kappa$ and $\lambda$ are $\mathcal{O}(10^{-2})$ and $\mathcal{O}(10^{2})$, which are fairly natural compared to the expectation of $\mathcal{O}(1)$. An enhanced SGWB spectrum can be achieved anywhere on the low-$\lambda$ edge of the allowed parameter space in Fig.~\ref{fig:3dparamspace}, so no great tuning is needed in any single parameter.

\section{Conclusions}
\label{sec:conclusions}

We have demonstrated, as a proof of concept, that a model which can produce a mild stiff period with slowly varying (increasing) barotropic parameter with \mbox{$\frac13<w(t)<1$} directly after inflation, such that the combination of mild stiff period and kination proper can be elongated, is able to overlap with the sensitivities of GW observations in the near future without disturbing the process of BBN. The model produces a distinctive $\Omega_{\textnormal{GW}}(f)$ spectrum with a rounded peak at high frequencies $f$ corresponding to the mild stiff period. We employed a toy model realisation of this idea, using a potential of exp(-exp) form, which can be motivated by string theory. We find that such a form of potential also has a pseudo-scaling attractor at large field values, which allows the field's energy density at late times to slowly decay away relative to the background.

We found that this model has a large parameter space in which the desired behaviour is observed, without requiring significant tuning to satisfy the observations. In Fig.~\ref{fig:GWspectrum}, we show an example point for which the peak of the GW spectrum is just short of the $\Delta N_{\textnormal{eff}}$ constraint and which overlaps with the observation contours of multiple upcoming experiments, especially with the observational capabilities of the Einstein Telescope (ET) and the Cosmic Explorer (CE).  
Observation of the shape of the spectrum of primordial gravitational waves can directly provide crucial information for the equation of state of the Universe post-inflation, during an era which so far has been extremely difficult to probe.

Our results can be compared to previous studies with features in $w_\phi$ and the SGWB spectrum. For example, a mild peak in the GW spectrum is obtained when, after inflation, the inflaton oscillates in a higher order potential minimum
\cite{Chen:2024roo}. Another example is studied
in Ref.~\cite{SanchezLopez:2023ixx}, where higher-order kinetic terms in the scalar field action produce a sharp `knee' in the spectrum. 
In Ref.~\cite{Dimopoulos:2022rdp}, a similar setup produces a gradual increase in $w_\phi$ towards kination, similarly to our model. Many more possibilities are investigated in Ref.~\cite{Gouttenoire:2021jhk}. However, our rounded peak is distinct from all these proposals. It would be interesting to study the SGWB signatures in these and other models further in the limit of a soft stiff period.

\acknowledgments
L.B. was supported by an STFC studentship. K.D. and E.T. were supported (in part) by the Consortium for Fundamental Physics under STFC consolidated grant:  ST/X000621/1.
E.T. was supported by the ``Fonds de la Recherche Scientifique'' (FNRS) under the IISN grant number 4.4517.08.

\bibliography{bibliography}

\end{document}